\documentstyle[12pt,aaspp4]{article}
\begin{document}

\newcommand{\up}[1]{\ifmmode^{\rm #1}\else$^{\rm #1}$\fi}
\newcommand{\zdot}{\makebox[0pt][l]{.}}
\newcommand{\upd}{\up{d}}
\newcommand{\uph}{\up{h}}
\newcommand{\upm}{\up{m}}
\newcommand{\ups}{\up{s}}
\newcommand{\arcd}{\ifmmode^{\circ}\else$^{\circ}$\fi}
\newcommand{\arcm}{\ifmmode{'}\else$'$\fi}
\newcommand{\arcs}{\ifmmode{''}\else$''$\fi}

\title{The Araucaria Project. An improved distance to the Sculptor spiral
galaxy NGC 300 from its Cepheid variables
}

\author{Wolfgang Gieren}
\affil{Universidad de Concepci{\'o}n, Departamento de Fisica, Astronomy Group, 
Casilla 160-C, 
Concepci{\'o}n, Chile}
\authoremail{wgieren@astro-udec.cl}
\author{Grzegorz Pietrzy{\'n}ski}
\affil{Universidad de Concepci{\'o}n, Departamento de Fisica, Astronomy Group, 
Casilla 160-C,
Concepci{\'o}n, Chile}
\affil{Warsaw University Observatory, Al. Ujazdowskie 4,00-478, Warsaw, Poland}
\authoremail{pietrzyn@hubble.cfm.udec.cl}
\author{Alistair Walker}
\affil{Cerro Tololo Interamerican Observatory, La Serena, Chile}
\authoremail{awalker@ctio.noao.edu}
\author{Fabio Bresolin}
\affil{Institute for Astronomy, University of Hawaii at Manoa, 2680 Woodlawn 
Drive, 
Honolulu HI 96822, USA}
\authoremail{bresolin@ifa.hawaii.edu}
\author{Dante Minniti}
\affil{Pontificia Universidad Cat\'olica de Chile, Departamento de Astronomia y 
Astrofisica,
Casilla 306, Santiago 22, Chile}
\authoremail{dante@astro.puc.cl}
\author{Rolf-Peter Kudritzki}
\affil{Institute for Astronomy, University of Hawaii at Manoa, 2680 Woodlawn 
Drive,
Honolulu HI 96822, USA}
\authoremail{kud@ifa.hawaii.edu}
\author{Andrzej Udalski}
\affil{Warsaw University Observatory, Aleje Ujazdowskie 4, PL-00-478, Warsaw, 
Poland}
\authoremail{udalski@astrouw.edu.pl}
\author{Igor Soszy{\'n}ski}
\affil{Universidad de Concepci{\'o}n, Departamento de Fisica, Astronomy Group, 
Casilla 160-C,
Concepci{\'o}n, Chile}
\affil{Warsaw University Observatory, Aleje Ujazdowskie 4, PL-00-478, Warsaw, 
Poland}
\authoremail{soszynsk@astrouw.edu.pl}
\author{Pascal Fouqu\'e}
\affil{Observatoire Midi-Pyr\'en\'ees, UMR 5572, 14 avenue Edouard Belin, 
F-31400 Toulouse, France}
\authoremail{pfouque@ast.obs-mip.fr}
\author{Jesper Storm}
\affil{Astrophysikalisches Institut Potsdam, An der Sternwarte 16, D-14482 
Potsdam, Germany}
\authoremail{jstorm@aip.de}
\author{Giuseppe Bono}
\affil{Osservatorio Astronomico di Roma, Via Frascati 33, 00040 Monte Porzio 
Catone, Italy}
\authoremail{bono@mporzio.astro.it}
\begin{abstract}
In a previous paper, we reported on the discovery of more than a hundred new 
Cepheid variables 
in the Sculptor Group
spiral NGC 300 from wide-field images taken in the B and V photometric bands at 
ESO/La Silla. In this paper,
we present additional VI data, derive improved periods and mean magnitudes for 
the variables, and
construct period-luminosity relations in the V, I and the reddening-independent 
(V-I) Wesenheit bands
using 58 Cepheid variables with periods between 11 and 90 days. We obtain 
tightly defined relations,
and by fitting the slopes determined for the LMC Cepheids by the OGLE II Project 
we obtain
reddening-corrected distances to the galaxy in all bands which show a slight 
offset to each other in the
sense that the Wesenheit relation yields the smallest distance, whereas the I- 
and V-band distances
are by 0.094 mag and 0.155 mag, respectively, larger. We adopt as our best value 
the distance derived from
the reddening-free Wesenheit magnitudes, which is 26.43 $\pm$ 0.04 (random) 
$\pm$
0.05 (systematic) mag.
The distance moduli from both, the V- and I-bands agree perfectly with the 
Wesenheit value if one
assumes an additional reddening of E(B-V)=0.05 mag intrinsic to NGC 300, in 
addition to the Galactic
foreground reddening towards NGC 300 of 0.025 mag. Such a modest intrinsic 
reddening is 
supported by recent HST images of NGC 300 which show that this galaxy is 
relatively dust-free, but also reveal 
that there must be {\it some} dust absorption in NGC 300. We argue that our 
current
distance result for NGC 300 is the most accurate which has so far been obtained 
using Cepheid variables,
and that it is largely free from systematic effects due to metallicity, 
blending, and sample selection.
It agrees very well with the recent distance determination from the tip of the 
red giant branch method
obtained from HST data by Butler et al. (2004), and it is consistent with the 
Cepheid distance to
NGC 300 which was derived by Freedman et al. (2001) from CCD photometry of a 
smaller sample of stars.

\end{abstract}

\section{Introduction}

In a previous paper (Pietrzy{\'n}ski et al. 2002; hereafter Paper I), we have 
reported
 on the discovery of 117 Cepheids
and 12 Cepheid candidates in the Sculptor Group spiral galaxy NGC 300 from B and 
V images obtained with the
 Wide Field Imager at the 2.2m telescope at ESO/La Silla, which strongly 
expanded the number of 18
 Cepheids in this galaxy which were known from the earlier pioneering work of 
Graham (1984), and
  Freedman et al. (1992).
  These data have been obtained as part of our
 ongoing Araucaria Project which seeks to provide improved calibrations of the 
dependences of several stellar
 distance indicators, including Cepheids, RR Lyrae stars, red clump giants and 
blue supergiants,
  on the environmental properties of their host galaxies (Gieren et al. 2001; 
see also
{\it http://ifa.hawaii.edu/\~{}bresolin/Araucaria/index.html}).
  NGC 300 is one of the principal target galaxies of this project. Our search 
for additional Cepheid variables 
 in this galaxy was motivated by two goals: first, to derive a more accurate 
Cepheid-based distance to
 this nearby galaxy than what was possible from the relatively sparsely sampled 
light curves of
 little more than a dozen of Cepheids
 (Freedman et al. 2001), and this way provide the basis for
  an accurate 
 calibration of several secondary stellar standard candles; and second, to 
provide an
 independent empirical calibration of the effect of metallicity on Cepheid 
absolute magnitudes,
 which is very much under dispute at the present time (e.g. Fouqu{\'e} et al. 
2003;
 Storm et al. 2004). NGC 300 seems well suited for such a calibration, due to 
the relatively large
 metallicity gradient in its disk suggested by H II region work (e.g. Deharveng 
et al. 1988),
  and whose value
 has recently been improved by the detailed spectroscopic work of our group on 
blue supergiant stars
  in this
 galaxy which we have been observing with the ESO VLT (Bresolin et al. 2002; 
Urbaneja et al. 2003, 2004).
 The relatively strong radial variation of metallicity in the disk of NGC 300, 
combined with the fact
 that we have discovered Cepheids over a large range of galactocentric 
distances,
 should allow a precise determination of the metallicity effect on their 
absolute magnitudes.
 This investigation will be the subject of a forthcoming paper.
 
 While our previous ESO BV imaging data were extremely useful for the purpose of 
finding a large number
 of Cepheids in NGC 300, we needed additional I-band data for accurate distance 
work, to enable
 us to address the problem of reddening via the reddening-independent (V-I) 
Wesenheit magnitudes. Some
 I-band images we had obtained during our previous runs with the ESO Wide Field 
Imager turned out to suffer from
 a strong fringing pattern, which we found extremely difficult to correct for in 
a reliable way. We therefore
 decided to re-observe NGC 300 in the I (and V) bands using the mosaic cameras 
attached to
 the 4-m telescope at CTIO, and to the Warsaw 1.3-m telescope at Las Campanas, 
which are basically
 free from this problem. We report on these new data in section 2, and use them
 together with the data given in Paper I to derive improved periods for the 
Cepheids (feasible
 due to the much enlarged time baseline of the observations), and to
 construct improved V-band, and I-band light curves for the variables, from 
which the mean magnitudes
 are determined. In section 3, we present the period-luminosity relations 
derived from our data in the various
 photometric bands which form the basis for the determination of the distance to 
NGC 300. Our distance
 determination is presented and
 discussed in section 4. In section 5, we discuss the sources of uncertainty
 which affect it, and compare to other determinations of the distance to NGC 300 
which have been reported
 in the literature. Section 6 summarizes our conclusions.

\section{Observations, Reductions and Calibrations}
The new data presented in this paper were collected with the Warsaw 1.3-m 
telescope at Las Campanas Observatory
and the 4-m telescope at CTIO. Each telescope was equipped with 
a mosaic 8k $\times$ 8k detector, with fields of view of about 35 $\times$ 35 
arcmin and a
scale of about 0.25 arcsec/pix. For more  instrumental
details on these cameras, the reader is referred to the corresponding websites:
{\it http://www.astrouw.edu.pl/\~{}ogle/index.html} {\it 
http://www.ctio.noao.edu/mosaic}.

The main body of the new data
was obtained with the Warsaw telescope at Las Campanas, but the additional CTIO 
4-m observations
turned out to be very useful, particularly to improve the I-band light curves of 
some of the 
Cepheids in NGC 300. Our new observations started on 2002 October 5
and lasted until 2003 November 11, providing a time baseline of nearly 4.5 years 
of modern data, 
together with the previous ESO/WFI
observations reported in Paper I, for the determination of improved periods for 
the Cepheids
in NGC 300, and for other variables we have discovered in this galaxy 
(Mennickent et al. 2004;
Bresolin et al. 2004).
During the new 2002-2003 period, we secured 30 mosaic images each in the V and I 
bands, 
on different nights. The Las Campanas
 observations were not dithered to compensate for the (small) gaps between the 
individual CCDs of the mosaic camera;
 for this reason, we lost a few of the Cepheids
 presented in the Cepheid catalog of Paper I. Also, some other Cepheids happened 
to appear too close to the edges
 of the individual CCDs to be properly calibrated. For these two reasons, we 
could not obtain I-band
 light curves for about 20 percent
 of the total number of Cepheids we had previously discovered in NGC 300.
 This was, however, not a reason to worry because the total number of Cepheids 
left to construct
 the PL relations in the V and I bands in the galaxy is still high enough for 
very accurate results (see section 4).
  With few exceptions, the seeing during the new observations was around 1", and 
sometimes better. 
 Integration times were 900 s per image (in both I and V) with the Warsaw
 telescope, and 300 s per image with the CTIO 4-m telescope. 

 Preliminary reductions (i.e. debiasing and flatfielding) of the CTIO data were 
done
   with the IRAF\footnote{IRAF is distributed by the
National Optical Astronomy Observatories, which are operated by the
Association of Universities for Research in Astronomy, Inc., under cooperative
agreement with the NSF.} "mscred" package. Then, the PSF photometry was obtained 
for all 
stars
in the same manner as for our WFI data from the ESO/MPI 2.2 m telescope (Paper 
I).
The data from the 1.3 m Warsaw telescope were reduced with the OGLE III pipeline 
based 
on the image subtraction technique (Udalski 2003, Wo{\'z}niak 2000).  

To calibrate each of our two data sets onto the standard system we used the 
extensive sequence of secondary standard 
stars, distributed over almost the whole observed area, which was established by 
Pietrzy{\'n}ski, Udalski and Gieren (2002). 
 
 \section{Cepheid Magnitudes and Revised Periods}
 The new V data obtained at Las Campanas and CTIO were merged with the previous 
 ESO/WFI data reported in Paper I
 to construct improved V light curves for the Cepheids. Excellent agreement 
between the different datasets,
 and in particular between the abundant ESO/WFI and Warsaw 1.3m datasets,
  was found, with no evidence 
 of zero-point shifts between the datasets for individual Cepheids amounting to 
more than 0.03 mag.
 Generally, it was not possible to detect {\it any} evidence for a zero point
 offset between the different datasets, which could in principle occur if the 
zero point
 was varying over the respective fields of the cameras we used. From our data, 
we can limit
 the size of such an effect, if existent at all,
 to less than 0.03 mag in both filters. 
  The result of merging the V data are light curves of exquisite quality for 
most of the variables, especially for the
 brighter, longer-period Cepheids which carry the largest weight in the distance 
determination
 (see section 4).
 In this process, we were able to improve the periods for
 a substantial fraction of the Cepheids, by fitting together our own datasets 
now spanning 
 more than 4 years, and by combining
 them with the older CCD data of Freedman et al. (1992)
 for those Cepheids which had already been known and observed prior to our 
wide-field study. While
 for most of the Cepheids the periods derived in Paper I were basically 
confirmed, and only slight
 modifications were found necessary, for a few of the stars, marked in Table 1 
(our revised Cepheid catalog), 
 we found evidence for periods changing in time. Among these stars with evidence
 for a variable period are the variables cep002 (V12), and cep005 (V3). Both 
stars have  periods in excess of $50^d$ and therefore  masses in the 10-13 
$M_{\odot}$ range,
and we could be
 witnessing a continuous change in their pulsation periods caused by their 
relatively fast
 evolutionary migration across the Cepheid instability strip which takes about
 $10^5$ yrs for the second crossing, and only $\sim$ $10^3$ yrs for the third 
crossing
 (Bono et al. 2000). Future monitoring of these stars will allow us to determine
 the period changes more precisely to check on this interesting possibility.
 
 The I-band light curves were constructed from our new data, adopting the 
periods derived from
 the analysis of the V-band data. They are of excellent quality for most of the 
longer-period stars, but
 even for the shorter-period, fainter Cepheids in our sample they are of good
 quality for most of the stars. Only for the shortest-period Cepheids in our 
sample, less than about
 10 days, the I-band
 light curves become quite noisy, as expected at the combination of telescope 
size and integration
 times we were using in the acquisition of these data.
  In some cases, we combined the new I-band magnitudes with those of
 Freedman et al. for common stars (whenever this procedure led to an improved 
light curve from the combined
 datasets-this was not always the case) before calculating the mean magnitudes.
  We also note here that there was no evidence for
 any significant zero point offset between the Freedman et al. data and our own 
data, in neither
 of the two bands. We had demonstrated this already for the V-band light curves 
in Paper I.
  This adds to our confidence that our adopted photometric zero-points in both,
 V and I are accurate to $\sim$0.03 mag.
 
 Table 1 presents the catalog of Cepheid data for all the variables having both 
V- and I-band light curves
 from our observations, and pulsation periods longer than 10 days (64 stars). We 
do not include
 in Table 1 the Cepheids with periods less than 10 days because they were not 
used in the distance
 determination for reasons discussed in section 5.
 Finding charts for each of these variables were given in Paper I, as were their 
coordinates.
 We do not repeat the mean
 magnitudes in the B band for these objects-they were reported in Paper I, and 
have not changed because we did
 not re-observe the stars in the B filter. The entries in Table 1 are the 
revised periods,
 the revised epochs of maximum brightness (in HJD) in V, and the
 intensity mean magnitudes in V (derived by Fourier series fitting to our own
 datasets), 
 I (from our own new datasets reported in this paper, merged with the
 Freedman et al. (1992) data whenever this increased the light curve quality), 
and the reddening-free
 (V-I) Wesenheit
 magnitude $W_{I}$ (defined as $W_{I}$ = I - 1.55 (V-I); see Udalski et al. 
1999). 
 The formal uncertainties on the intensity mean magnitudes in all bands, as 
returned
 from the Fourier fitting routine, are always in the range 0.01-0.02 mag. Their 
small sizes 
 reflect the high quality of the light curves for most
 of the variables. The periods in Table 1 replace the values given in Paper I, 
as do the mean
 V magnitudes. To demonstrate the quality of our data, we show the V- and I-band 
light curves for some of the Cepheids
 of our sample in Fig. 1. These light curves are representative for the average 
quality
 of the light curves of other Cepheids in our sample of similar period. Fig. 2 
shows a Cepheid light
 curve in V and I in which we have distinguished the different datasets with 
different symbols,
 in order to demonstrate the very good agreement among them. Such a good 
agreement is seen for
 nearly all the Cepheids, albeit the noise in some light curves is larger.
 
 Table 2 contains the individual new V and I observations for all the Cepheids 
in Table 1. The full
 Table 2 is available in electronic form.

\section{The Distance Modulus of NGC 300}
From the data given in Table 1, we constructed the period-luminosity relations 
for the Cepheids in NGC 300
in V, I and $W_{I}$. We excluded the one star in our Table having a period 
longer than 100 days,
cep001. The reason is that at periods exceeding 100 days, previous evidence 
seems to indicate that the Cepheid 
PL relation might change its slope and become nearly flat (see, for instance, 
Fig. 4 in Freedman et al. 1992). Such an effect is also predicted from 
theoretical
models (Bono et al. 2002). 
Indeed, cep001 with its period of 115 days is more than half a magnitude fainter 
than what we would
expect from the PL relations defined by all the other stars. Besides this 
(probably) peculiar Cepheid,
we rejected 5 Cepheids due to their large deviations from the mean PL relations 
in one, or several
bands. These variables are marked in Table 1. They all have relatively short 
periods (where we have many
Cepheids, and their relative weight in the fits is therefore low).
We note, however, that their inclusion in the fits and distance solutions would 
introduce only
very minor, insignificant changes in the PL relations in the different bands, 
and in the resulting
distance to NGC 300. We chose a short-period cutoff period of 10 days for the 
construction of the PL relations
for several reasons. First, inspection of the PL relations in V and B which were 
constructed from
our previous, smaller dataset in Paper I (Fig. 10 there) shows that a Malmquist 
bias (Sandage 1988) is present for periods
shorter than about 10 days. This is due to the photometric detection limit in 
our data - we start to miss
the intrinsically faint Cepheids at periods less than about 10 days (the ones 
lying below the ridge
line in the Cepheid instability strip). For periods of 10 days or longer, we are 
obviously not limited by
this Malmquist bias anymore, and our stars can be expected to fill the Cepheid 
instability strip
in a homogeneous way.
A second reason to omit the very short-period stars in our Cepheid sample is the 
fact that these,
intrinsically relatively faint Cepheids are more susceptible to blending than 
the 
longer-period, brighter Cepheids.
We discuss the effect of blending on our distance 
results in more detail in section 5,
but remark here that using only the longer-period Cepheids for distance 
determinations will protect us,
to a larger extent, against the possible effect of unresolved companion stars on 
the solutions. 
The third reason to exclude Cepheids of period less than 10 days in our 
solutions is the fact
that just downwards from about this period, the light curves (particularly those 
in I which were
basically measured at a 1.3m telescope) become too noisy, increasing the random 
uncertainty in our
distance solution. 

In Figs. 3 and 4, we show the period-luminosity (PL) relations in the V and I 
bands obtained from the
58 Cepheids adopted for the final distance solutions.
These figures show that our data define a tight PL relation in the V band, and 
an
even tighter relation in the I band, which is expected due to the smaller 
intrinsic width of the Cepheid
instability strip toward longer wavelengths, and to the reduced influence of a 
possible differential
extinction inside the galaxy. It is seen that there is no evidence for curvature 
in the relations
at the shortest periods, indicating that our choice for the value of the cutoff 
period is adequate.
We de-reddened our V and I mean magnitudes adopting a Galactic foreground 
reddening to NGC 300
of E(B-V)=0.025 mag (Burstein \& Heiles 1984), and an extinction law of 
$A_{V}$=3.24 E(B-V), and
$A_{I}$=1.96 E(B-V) (Schlegel et al. 1998). There might of course exist some 
additional
absorption produced inside NGC 300; however, there are reasons to believe that 
such an intrinsic
absorption is small (see discussion in Freedman et al. 1992), which seems to be 
confirmed by
the recent high-resolution image obtained for NGC 300 with HST and which has 
been published in the
Hubble Heritage Project, showing that NGC 300 is indeed relatively free of dust.
In any case, our adopted approach for eliminating any influence of reddening
on our results, or at least reduce it to the lowest possible level, is to use 
the reddening-free Wesenheit PL
relation for the distance solution. We display the $W_{I}$-log P relation from 
our data in Fig. 5.
Again, the data define a relationship which is evidently linear over the whole 
period range from 10-90 days we use, 
and whose dispersion is, as expected, somewhat smaller than the one of the 
I-band PL relation.

In order to derive the NGC 300 distance from the Wesenheit PL relation, we 
decided to adopt the slope
for this relation as established by Udalski (2000) from the LMC Cepheids. There 
are two reasons for
this choice. First, the slopes and zero points of the Cepheid PL relations in 
the LMC in W, as well as in V and I,
are extremely well established from the OGLE II Project (Udalski et al. 1999; 
Udalski 2000). Second,
there is some evidence that the slope of the Cepheid PL relation might depend, 
to some degree,
on metallicity (e.g. Storm et al. 2004; Tammann et al. 2003; for a more detailed 
discussion on this point,
see section 5). To minimize any possible systematic effect of metallicity on our 
distance determination
for NGC 300, especially in view of the current uncertainty of the effect a 
changing metallicity has on
Cepheid absolute magnitudes, it seems wise to use the PL slopes derived in a 
galaxy which has a similar
mean metallicity as the target galaxy under study. From our recent spectroscopic 
work on blue
supergiant stars in NGC 300 (Urbaneja et al. 2003, 2004) 
we now know that the mean metallicity of these stars, which are quite
evenly distributed over the disk of NGC 300, is about -0.3 to -0.4 dex, which is 
practically
identical to the mean LMC metallicity for the young stellar population, 
including Cepheids (Luck
et al. 1998). Therefore it is, in the case of NGC 300, certainly the best choice 
to adopt the
PL relation slopes determined by the OGLE II Project in the LMC. Nevertheless, 
we also did "free fits"
to the data, without pre-specifying the slopes. These yield values of -3.01 
$\pm$ 
0.16, -3.12 $\pm$ 0.12,
and -3.29 $\pm$ 0.11 for the PL relation slopes in V, I  and $W_{I}$, 
respectively. These values 
compare very well with the corresponding OGLE II LMC slopes of -2.775, -2.977 
and -3.300, supporting
our choice.
Particularly in the Wesenheit band, the agreement of the NGC 300 PL relation 
slope with that defined by
the LMC Cepheids is striking-the values are virtually identical. The agreement 
of the slopes
determined from the NGC 300 Cepheids with the ones obtained from Galactic 
Cepheids whose distances
have been measured with the near-infrared surface brightness technique of 
Fouqu{\'e} \& Gieren (1997)
(see Gieren, Fouqu{\'e} \& G{\'o}mez 1998; Storm et al. 2004) is not as good, 
the Galactic Cepheid
slopes being steeper than the slopes we observe in NGC 300, particularly in the 
Wesenheit band
which is the most important one for the distance determination. We suspect that 
this difference is
caused by the higher average metallicity of the Galactic Cepheid sample.

Using the OGLE II slopes given above, we derive the following equations for the 
PL relations in NGC 300:\\

          $V_0$ = -2.775 log P + (25.155 $\pm$ 0.034)   \hspace*{1cm} rms=0.275  
mag  \\
          
          $I_0$ = -2.977 log P + (24.621 $\pm$ 0.025)    \hspace*{1cm}     
rms=0.205 mag  \\    
 
          $W_{I}$ = -3.300 log P + (23.802 $\pm$ 0.028)   \hspace*{1cm}    
rms=0.198  mag  \\

The rms dispersions around these mean relations are somewhat larger, but quite 
comparable, to
those for the Galactic Cepheid PL relations (Storm et al. 2004), which seems 
quite remarkable for
such a faint and distant sample of stars, and reinforces the very good quality 
of our data.

Combining the OGLE II LMC PL relation zero points (Udalski, 2000) in the three 
photometric bands
 with those of the previous equations,
we obtain the distance of NGC 300, {\it with respect to the LMC}, directly by 
taking the difference
(ZP(NGC 300) - ZP(OGLE II). This can be transformed to an {\it absolute} 
distance to NGC 300 by adding
the adopted LMC distance modulus. Without going into the difficult discussion 
about the true value of the
LMC distance modulus here (see Benedict et al. 2002; Walker 2003; Feast 2003), 
we adopt as the
current best distance modulus to the LMC the value 18.50. With this adopted 
distance to the LMC,
we find the following distance moduli for NGC 300, in the 3 different bands we 
study:

          $(m-M)_0$ ($W_{I}$) = 26.434 mag
          
          $(m-M)_0$ (I) = 26.528 mag
          
          $(m-M)_0$ (V) = 26.589 mag

Whereas the values from the 3 different bands agree quite closely among each 
other, there {\it is} some scatter
beyond the random uncertainties on these numbers (see section 5).
 We adopt the
distance coming from the reddening-free Wesenheit index as our best result. This 
is also supported by
the fact that the Wesenheit relation shows the smallest scatter of the three 
relations, and its observed slope
is in truly perfect agreement with the slope defined by the LMC Cepheids. 

Why do the I- and V-band data yield slighly larger distance moduli? A 
straightforward and very reasonable
explanation is that our adopted Galactic foreground absorption is insufficient, 
and that we have neglected
some contribution to E(B-V) intrinsic to NGC 300. Indeed, assuming an average 
{\it intrinsic} (to
NGC 300) reddening of E(B-V)=0.05 mag, in addition to the foreground 
E(B-V)=0.025 mag, the
distance moduli derived from {\it both}, I and V, converge, within 0.01-0.02 
mag, to our adopted
NGC 300 distance modulus from the Wesenheit PL relation, and we have a 
completely consistent solution
from the three bands. 

A possible concern in our way to compare the samples of Cepheids in the LMC
and NGC 300 is the fact that the mean periods of the two samples are quite 
different,
the mean period of the LMC sample observed by the OGLE II Project being much 
shorter.
It is therefore interesting to repeat the above analysis, for comparison, on the 
subsample
 of OGLE II LMC Cepheids
having periods larger than 10 days, just as our adopted NGC 300 sample. 
Recently, Kanbur et al. (2003)
has identified some long-period Cepheids in the OGLE sample which have problems 
like
strange shapes of their light curves, large phase gaps in their light curves, or 
unusually
low light curve amplitudes. Taking out such stars from the sample,
we get a "clean" sample of 45 LMC Cepheids with periods in excess of 10 days. If 
we
fit the same slopes as given above (obtained by Udalski (2000) from the full 
sample of
Cepheids of all periods) to the PL relations defined by these 45 long-period LMC 
Cepheids, and
compare the resulting zero points (very slightly different from the zero points 
coming
from the full sample of LMC Cepheids) to the NGC 300 zero points in the 
different bands,
we get almost {\it exactly} the same distance to NGC 300 from the Wesenheit 
magnitudes
(26.444 mag, instead of 26.434 mag). The offsets of the distances now obtained 
in the V and I
bands to the W band are slightly smaller than above, and are consistent with an
additional internal reddening in NGC 300 of 0.03 mag, as compared to the 0.05 
mag
we derived above. It is therefore evident that our derived distance to NGC 300 
does not
depend on introducing a period cutoff of 10 days in the LMC sample, in order to 
make its mean period
comparable to our NGC 300 sample, or just working with the complete sample of 
LMC
Cepheids of all periods.

\section{Discussion}
In this section, we are discussing the sources of error which may affect our 
distance result, and
will try to estimate their effect as realistically as possible. We will not 
discuss the current uncertainty
of the LMC distance modulus (which clearly remains the largest source of 
systematic error in
the adopted absolute distance to NGC 300)-regarding this problem, the reader is 
referred to the papers
of Walker (2003), Feast (2003), and Benedict et al. (2002), and references given 
therein.
\subsection{Photometry}
An obvious source of systematic uncertainty on our distance result is in the 
adopted zero points
of our photometry. We already mentioned that our different datasets are very 
consistent among
themselves, excluding systematic zero point variations among the datasets 
exceeding ~0.03 mag, both in
V and I. This is of course not surprising because we used identical standard 
stars and reduction procedures
for all our datasets. The comparison with the Freedman et al. (1992) photometry 
for the Cepheids
in common supports the conclusion that zero point errors are not exceeding 0.03 
mag. We already
discussed the possibility that in addition to a systematic zero point error in 
our photometry,
there could be a slight zero point variation over the mosaic detectors
we used. A comparison of our different datasets for all the Cepheids
in our database (which appear in very different positions on the detectors) 
convinced us that such
an effect, if at all present, must be very small, certainly less than 0.03 mag. 
In any case, such
an effect would not be systematic, but just add to the random scatter in our 
derived mean magnitudes
of the Cepheids.

\subsection{Sample Selection for the Construction of the PL Relations}
In the previous section, we have already discussed the reason for adopting a 
period cutoff of 10 days
for our sample. The introduction of this cutoff effectively eliminates any 
significant contribution
to the systematic error on our NGC 300 distance result from a Malmquist bias due 
to the Cepheid
detection limit set by our photometry. It assures that the filling of the 
Cepheid instability strip
in the period range adopted for our study is homogeneous, and near-complete, 
given the large number
of Cepheids (58) we were able to use in our study. It also eliminates any 
possibility that our sample
is contaminated by the inclusion of overtone Cepheids. These stars do only 
appear at periods
smaller than about 8 days, as has been impressingly demonstrated by the 
different microlensing
projects (OGLE, MACHO, EROS) in the LMC.

In order to investigate if there is any dependence of our distance result on the 
adopted value of
the cutoff period, we also did solutions using period cutoffs of log P=1.2, 1.4 
and 1.5. In all
cases, the distance results were the same within 0.03 mag, but the uncertainty 
in the zero point
of the solutions increases as a result of having less stars for the fits. We can 
therefore be sure
that our result is very robust regarding the choice of the cutoff period, as 
long as this is large enough
to eliminate the problems due to Malmquist bias.

\subsection{Adopted Slopes of the Period-Luminosity Relations}
In the previous section, we already discussed the reason for adopting the slopes 
derived for the LMC
Cepheids by the OGLE II team. We believe that these reasons are very strong.
A reason of concern in using them for our current study might be the fact that
the OGLE II slopes are mostly based on Cepheids having periods smaller than 10 
days
while we use only Cepheids with periods larger than 10 days in NGC 300, but we 
have
shown in the previous section that this has no significant effect on our
distance solution.

Several of the
authors of this paper have made a great effort in the past to calibrate the 
{\it Galactic} Cepheid PL
relation with a high accuracy (Gieren et al. 1998; Storm et al. 2004). Yet, the 
current precision we
have achieved from the measurement of the distances of some 40 Galactic Cepheids
 cannot compete with the results for the LMC coming from 
several hundreds of stars.
The Galactic PL relation, if future work confirms that it is significantly 
different from the
LMC relation, will be useful in the study of galaxies which have near-solar 
metallicity. In the case
of NGC 300, it seems definitively preferable, from all the points of view we 
mentioned, to
use the OGLE LMC PL relations to derive an accurate, albeit relative, distance 
for this galaxy, which
can be scaled to any LMC distance which might be adopted in the future, in the 
light of new results. 

\subsection{Metallicity Effects}
The question of how chemical abundances affect the absolute magnitudes of 
Cepheid variables, and
how such a possible metallicity dependence translates into a shift in the zero 
points of the PL
relations in different photometric bands, has been the subject of a longstanding 
debate. Yet,
a clear answer to this question is still missing. All the observational tests 
which have been
carried out over the years (for a recent review, see Fouqu{\'e} et al. 2003; see 
also the recent
paper of Sakai et al. 2004) have suffered from the low accuracy of the reported 
results; therefore, the
metallicity effect remains largely unconstrained, at the present time. At least, 
evidence seems 
to be hard enough now to determine the sign of the effect: more metal-poor 
Cepheids are, at any
given period, intrinsically fainter than their more metal-rich counterparts, at 
least at visible
wavelengths, but by which amount is very uncertain. Clearly, more stringent 
tests will have to be
made, and it has been one of the principal motivations for our group to discover 
Cepheid variables 
in NGC 300 to carry out such an improved test. For the time being, it is clearly 
the best strategy
to minimize metallicity-related effects in Cepheid-based distance determinations 
by comparing
samples of Cepheids having the same average metallicity. Given that the mean 
metallicity of NGC 300
is very close to that of the LMC (references cited before), we are quite 
confident that there
is no significant contribution to our error budget from any metallicity-related 
effect. The expected variation in metallicity among the Cepheids in our
sample (a few tenths of 1 dex) could introduce some additional dispersion in our 
PL
relations, but such an effect is obviously quite small (see section 5.6.).

\subsection{Reddening}
In our approach, we have circumvented the problem of applying appropriate 
reddening corrections by
using the reddening-independent Wesenheit magnitudes. The relatively low scatter 
in the $W_{I}$-log P
relation in Fig. 4 demonstrates that the random errors in the mean magnitudes of 
the Cepheids in V and I
are low enough to successfully apply this method. The fact that we do not see a 
{\it dramatic} reduction
in the scatter of this relation, as compared to the I-band PL relation, is 
likely to be a consequence of the fact
that the reddening, including possible differential reddening inside NGC 300, is 
evidently small in the 
case of NGC 300, which already helps to reduce the scatter in the 
reddening-dependent V- and I-band PL relations.
The fact that a small additional reddening of 0.05 makes our distance results
from V, I and W fully consistent lends additional support to our conclusion that 
our distance
result is not significantly biased due to an unproper treatment of reddening.

Just to stress the relative independence of our distance result for NGC 300 on 
reddening,
we mention that if we adopt a constant reddening of E(B-V)=0.10 mag for the 
Cepheids in the LMC as
done by Fouqu{\'e} et al. 2003 (instead of the variable reddening from the OGLE 
maps),
and use the resulting very slightly changed OGLE II PL relations for the LMC,
 we obtain a Wesenheit distance of 26.40 mag for NGC 300, instead of
our adopted value of 26.43 mag. The difference is clearly not significant and 
reinforces
our conclusion that the effect of reddening on our result is very small.   

\subsection{Crowding Effects}
At the distance of NGC 300, crowding is clearly an issue at the typical 
resolution of 1 arcsec of our images.
Probably to each of the variables there are a few very nearby stars which are 
not resolved in the images
and therefore contribute to the measured fluxes of the Cepheids. The effect is 
systematic because it
makes the Cepheids too {\it bright}, and therefore, if the effect is 
significant, makes the derived
distance too small. While it is quite difficult to quantify the effect blending 
may have on the 
distance we derive for NGC 300, there are several indications that the effect is 
very small. One
reason comes from the observed relatively small rms scatter in all bands, which 
can be explained just 
by the random scatter in the mean magnitudes due to the quality of our 
photometry, the intrinsic width
of the instability strip, some contribution from differential reddening in the 
V- and I-band PL relations,
and possibly a small metallicity-related term due to the fact that the Cepheids 
in our sample will
have a dispersion in their individual metallicities of a few tenths of a dex. If 
a significant fraction
of the Cepheids would be heavily blended such that their observed fluxes are 
significantly altered,
they should stand out towards the bright end in the PL relations, and this 
should increase the
rms dispersion in the relations. There is very little evidence for this to occur 
in our data. Also, we
have protected ourselves against such an effect by eliminating those Cepheids 
which show a very large
deviation from the mean relations, and three of the five eliminated objects are 
indeed very bright
(see Figs. 3-5) and could be stars blended with exceptionally bright nearby 
companions. A second reason
why we think that our distance is not significantly affected by the blending 
problem is that we do not
see any indication of a flattening of the PL slopes towards the short-period end 
in the PL relations. It
can be reasonably assumed that blending will, on average, affect the 
short-period Cepheids more strongly 
than the longer-period ones because they are intrinsically fainter, by up to 3 
magnitudes. If 
blending is a serious issue, we should then expect a higher fraction of 
overluminous Cepheids close to our short
period cutoff than at larger periods. This should flatten the slope towards 
short periods (just as the
Malmquist bias, but for a totally different reason).
 We tested for this by adopting successively larger cutoff
periods up to log P=1.4; in all cases, the slopes of the resulting PL relations 
do not show
any evidence for a significant increase. For the arguments given, we believe 
that our distance determination
is not significantly affected by the problem of unresolved nearby stars. 
Fortunately, Cepheids are
intrinsically very bright stars and therefore their susceptibility to blending 
is small. 

Summarizing the previous discussion, our conclusion is that we have a total 
random error of ~0.04 mag
from the photometric noise, and a possible slight zero point variation over the 
CCD, and a total
systematic uncertainty which amounts to ~0.05 mag, and whose main contributor is 
the uncertainty
of our adopted overall photometric zero points in I and V. 
We therefore find a final result for
the distance to NGC 300 of

          $(m-M)_0$ = 26.43 $\pm$ 0.04 (random) $\pm$ 0.05 (systematic) mag
          
This distance result compares very favorably with the previous determination of 
Freedman et al. (2001)
who obtained a true distance modulus of 26.53, adopting the same value (18.50) 
for the LMC distance
modulus, and the same PL relation slopes from the OGLE II Project. It is 
consistent with this value within
the combined 1 $\sigma$ uncertainties. It is also clearly consistent, within the 
combined uncertainties, 
with another recent HST-based determination of the distance to NGC 300 from the 
I-band magnitude of the
tip of the red giant branch, which yielded a distance modulus of 26.56 $\pm$ 
0.07 
($\pm$ 0.13)mag
(Butler et al. 2004).

\section{Conclusions}
We have used a large number of wide-field images in the V and I bands which have 
been obtained 
over almost four years at ESO, Las Campanas and CTIO to discover a large number 
of Cepheid variables
in the Sculptor spiral galaxy NGC 300. Our images have been carefully analyzed 
and calibrated,
and we have used the long-period Cepheids in the galaxy ($P\geq 10 days$) to 
establish 
the period-luminosity
relations in the V, I and $W_{I}$ bands. The definition of these relations from 
our data
is excellent, and their slopes agree very well with those found by Udalski et 
al. (1999) for the
Cepheids in the LMC. In order to minimize metallicity-related effects on our 
distance determination,
we used the LMC slopes to fit our NGC 300 sample, given that the mean 
metallicity for both galaxies
appears to be the same. The fit to the reddening-independent Wesenheit PL 
relation yields our
best, adopted distance to NGC 300 of 26.43 mag, with small statistical (0.04 
mag) and systematic
(0.05 mag) uncertainties. This value assumes that the LMC distance is 18.50 mag, 
but
can be scaled to any different value of the LMC distance if this turns out to be 
appropriate
in the light of future work. We find somewhat longer distances from the V- and 
I-band PL relations assuming
a Galactic foreground reddening of E(B-V)=0.025; if we postulate an additional 
reddening of 0.05, intrinsic
to NGC 300, the distances obtained from the V and I bands agree perfectly with 
that derived
from the reddening-independent W band. The distance derived in this paper is the 
most accurate one
so far measured to NGC 300 from Cepheid variables, but it agrees well with the 
previous determination
of Freedman et al. which was based on CCD photometry of a much smaller sample of 
variables. Our
distance determination is also consistent with the recent determination of 
Butler et al. from 
the TRGB method, using I-band images obtained with the Hubble Space Telescope. 
The distance to
NGC 300 seems therefore to be well established now. To confirm this, and 
hopefully further improve the
accuracy of our current result, we will use near-infrared Cepheid PL relations 
in a study which
will be carried out soon. NGC 300 will then be a key galaxy to calibrate 
secondary stellar methods
of distance determination, as we are proposing to do in our Araucaria Project.

\acknowledgments
We are grateful to the European Southern Observatory, to Las Campanas 
Observatory and to the
Cerro Tololo Inter-American Observatory for providing the large amounts of 
telescope time
which were necessary to complete this project. We also very much appreciate the 
support
we received from the staffs of all these institutions.
WG, GP and DM gratefully acknowledge 
financial support for this
work from the Chilean Center for Astrophysics FONDAP 15010003. 
WG also acknowledges support
from the Centrum fuer Internationale Migration und Entwicklung in
Frankfurt/Germany
who donated the Sun Ultra 60 workstation on which a substantial part of the data 
reduction and
analysis for this project was carried out.   
Support from the Polish KBN grant No 2P03D02123 and BST grant for 
Warsaw University Observatory is also acknowledged.

\begin{deluxetable}{c c c c c c c c}
\tablecaption{Cepheids in NGC 300}
\tablehead{
\colhead{ID} & \colhead{P [days]} & \colhead{log P} & \colhead{ ${\rm 
T}_{0}-2450000$} & 
\colhead{$<V>$} & \colhead{$<I>$} & \colhead{$<W_{\rm I}>$} & \colhead{Remarks}
}

\startdata
cep001 &  115.8 &   2.0637 & 2952.5 &   20.134 &  19.162 &   17.655 &rejected from the fit\\ 
cep002 &  89.06 &   1.9497 & 2911.65 &   19.712 &  18.696 &   17.121 &period variable ?\\ 
cep003 &  83.00 &   1.9191 & 1543.57 &   19.258 &  18.490 &   17.300 &\\ 
cep004 &  75    &   1.8751 & 2607 &   19.776 &  19.029 &   17.871 &\\ 
cep005 &  56.599 &   1.7528 & 1391.862 &   20.412 &  19.599 &   18.339 &period variable ?\\ 
cep006 &  52.751 &   1.7222 & 2965.566 &   20.465 &  19.549 &   18.129 &\\ 
cep007 &  43.35  &   1.6370 & 1547.60 &   20.917 &  19.917 &   18.367 &\\ 
cep008 &  40.29 &   1.6052 & 2853.85 &   20.317 &  19.586 &   18.453 &\\ 
cep009 &  36.74 &   1.5651 & 2871.77 &   21.068 &  20.220 &   18.906 &\\ 
cep010 &  35.58 &   1.5512 & 2907.66 &   21.264 &  20.210 &   18.576 &\\ 
cep012 &  34.999 &   1.5439 & 2899.708 &   20.869 &  20.015 &   18.691 &\\ 
cep013 &  34.75 &   1.5410 & 1408.85 &   20.828 &  19.893 &   18.444 &\\ 
cep014 &  33.975 &   1.5312 & 2928.635 &   20.701 &  19.886 &   18.623 &\\ 
cep015 &  32.29 &   1.5091 & 2859.79 &   21.006 &  20.155 &   18.836 &\\ 
cep016 &  28.41 &   1.4535 & 2853.85 &   21.339 &  20.275 &   18.626 &\\ 
cep018 &  25.003 &   1.3980 & 1488.572 &   21.446 &  20.666 &   19.457 &\\ 
cep019 &  24.79 &   1.3943 & 2911.65 &   21.583 &  20.703 &   19.339 &\\ 
cep022 &  24.24 &   1.3845 & 2849.81 &   21.609 &  20.828 &   19.617 &\\ 
cep023 &  24.037 &   1.3809 & 2849.819 &   21.893 &  20.903 &   19.369 &\\ 
cep026 &  23.444 &   1.3700 & 2903.740 &   21.227 &  20.426 &   19.184 &\\ 
cep027 &  23.35 &   1.3683 & 2911.65 &   21.271 &  20.369 &   18.971 &\\ 
cep028 &  23.12 &   1.3640 & 2916.65 &   20.992 &  20.341 &   19.332 &\\ 
cep029 &  22.79 &   1.3577 & 1519.58 &   21.808 &  20.786 &   19.202 &\\ 
cep030 &  22.21 &   1.3465 & 2907.66 &   21.533 &  20.618 &   19.200 &\\ 
cep032 &  21.07 &   1.3237 & 2856.85 &   21.424 &  20.700 &   19.578 &\\ 
cep035 &  19.485 &   1.2897 & 2916.658 &   21.460 &  20.733 &   19.606 &\\ 
cep036 &  18.91 &   1.2767 & 1519.59 &   22.469 &  21.406 &   19.758 &\\ 
cep038 &  18.24 &   1.2610 & 2552.53 &   21.543 &  20.873 &   19.834 &\\ 
cep039 &  18.31 &   1.2627 & 1438.84 &   22.165 &  21.216 &   19.745 &\\ 
cep040 &  18.219 &   1.2605 & 1484.641 &   21.478 &  20.745 &   19.609 &\\ 
cep041 &  18.012 &   1.2556 & 2952.562 &   21.441 &  20.751 &   19.681 &\\ 
cep043 &  17.833 &   1.2512 & 2552.540 &   21.515 &  20.804 &   19.702 &\\ 
cep044 &  17.22 &   1.2360 & 2898.70 &   22.055 &  21.275 &   20.066 &\\ 
\enddata
\end{deluxetable}

\setcounter{table}{0}

\begin{deluxetable}{c c c c c c c c}
\tablecaption{Cepheids in NGC 300 - continued}
\tablehead{
\colhead{ID} & \colhead{P [days]} & \colhead{log P} & \colhead{${\rm 
T}_{0}-2450000$} 
& \colhead{$<V>$} & \colhead{$<I>$} & \colhead{$<W_{\rm I}>$} & 
\colhead{Remarks}
}
\startdata
cep045 &  16.92 &   1.2284 & 1438.84 &   21.664 &  20.838 &   19.558 &\\ 
cep046 &  16.56 &   1.2191 & 2860.80 &   22.499 &  21.551 &   20.082 &\\ 
cep048 &  16.49 &   1.2172 & 2851.79 &   21.623 &  20.987 &   20.001 &rejected from the fit\\ 
cep049 &  16.10 &   1.2068 & 2849.81 &   22.085 &  20.951 &   19.193 &\\ 
cep050 &  15.92 &   1.2019 & 2928.63 &   21.762 &  21.005 &   19.832 &\\ 
cep051 &  15.70 &   1.1959 & 2886.69 &   22.053 &  21.248 &   20.000 &\\ 
cep052 &  15.61 &   1.1934 & 2879.76 &   21.575 &  20.813 &   19.632 &\\ 
cep053 &  15.49 &   1.1901 & 2907.66 &   22.225 &  21.375 &   20.057 &\\ 
cep055 &  15.0 &   1.1761 & 2911.6 &   22.190 &  20.994 &   19.140 &rejected from the fit\\ 
cep056 &  15.05 &   1.1775 & 1516.62 &   21.666 &  21.183 &   20.434 &rejected from the fit \\ 
cep058 &  14.80 &   1.1703 & 2907.66 &   22.039 &  21.123 &   19.703 &\\ 
cep059 &  14.56 &   1.1632 & 1488.57 &   22.117 &  21.455 &   20.429 &rejected from the fit\\ 
cep060 &  14.411 &   1.1587 & 1491.590 &   22.149 &  21.347 &   20.104 &\\ 
cep061 &  14.24 &   1.1535 & 2851.79 &   22.299 &  21.421 &   20.060 &\\ 
cep062 &  14.355 &   1.1570 & 1431.835 &   22.075 &  21.097 &   19.581 &\\ 
cep063 &  14.361 &   1.1572 & 2879.768 &   21.952 &  21.102 &   19.784 &\\ 
cep065 &  14.045 &   1.1475 & 2879.768 &   21.848 &  21.091 &   19.918 &\\ 
cep066 &  13.856 &   1.1416 & 2911.659 &   22.046 &  21.205 &   19.901 &\\ 
cep067 &  13.754 &   1.1384 & 2903.740 &   22.338 &  21.524 &   20.262 &\\ 
cep069 &  13.605 &   1.1337 & 1547.610 &   22.325 &  21.367 &   19.882 &\\ 
cep070 &  13.523 &   1.1311 & 2849.819 &   22.027 &  21.292 &   20.153 &\\ 
cep071 &  13.443 &   1.1285 & 2641.660 &   21.893 &  21.217 &   20.169 &\\ 
cep072 &  13.505 &   1.1305 & 2946.688 &   22.340 &  21.280 &   19.637 &\\ 
cep073 &  13.464 &   1.1292 & 1431.837 &   21.881 &  21.067 &   19.805 &\\ 
cep074 &  13.358 &   1.1257 & 1433.828 &   21.930 &  21.198 &   20.063 &\\ 
cep076 &  13.168 &   1.1195 & 1488.572 &   21.857 &  20.901 &   19.419 &rejected from the fit \\ 
cep077 &  12.94 &   1.1119 & 2916.65 &   22.230 &  21.430 &   20.190 &\\ 
cep079 &  11.951 &   1.0774 & 2911.659 &   22.053 &  21.298 &   20.128 &\\ 
cep081 &  11.579 &   1.0637 & 2911.659 &   22.287 &  21.582 &   20.489 &\\ 
cep082 &  11.497 &   1.0606 & 2903.740 &   22.635 &  21.765 &   20.416 &\\ 
cep085 &  11.2 &   1.0492 & 1524.6 &   22.208 &  21.457 &   20.293 &\\ 
\enddata

\end{deluxetable}

\begin{deluxetable}{ccccc}
\tablecaption{Individual New V and I Observations}
\tablehead{
\colhead{object}  & \colhead{filter} &
\colhead{HJD-2450000}  & \colhead{mag}  & \colhead{$\sigma_{mag}$}\\
}
\startdata
cep001 & V  &  2641.660400 &  19.734 &   0.013  \\ 
cep001 & V  &  2849.819092 &  20.670 &   0.024  \\ 
cep001 & V  &  2851.790527 &  20.634 &   0.025  \\ 
cep001 & V  &  2856.855957 &  20.527 &   0.026  \\ 
cep001 & V  &  2857.820557 &  20.507 &   0.021  \\ 
cep001 & V  &  2859.794189 &  20.489 &   0.031  \\ 
cep001 & V  &  2860.801270 &  20.444 &   0.032  \\ 
cep001 & V  &  2861.866699 &  20.360 &   0.036  \\ 
cep001 & V  &  2863.755127 &  20.303 &   0.058  \\ 
cep001 & V  &  2871.775635 &  19.868 &   0.014  \\ 
cep001 & V  &  2879.768311 &  19.633 &   0.011  \\ 
cep001 & V  &  2886.691406 &  19.801 &   0.019  \\ 
cep001 & V  &  2893.736572 &  19.907 &   0.039  \\ 
cep001 & V  &  2898.709229 &  19.916 &   0.019  \\ 
cep001 & V  &  2899.708008 &  19.911 &   0.014  \\ 
cep001 & V  &  2903.740479 &  19.989 &   0.012  \\ 
cep001 & V  &  2907.663086 &  20.001 &   0.015  \\ 
cep001 & V  &  2911.659668 &  20.010 &   0.013  \\ 
cep001 & V  &  2916.658203 &  20.076 &   0.021  \\ 
cep001 & V  &  2922.662354 &  20.186 &   0.053  \\ 
cep001 & V  &  2928.635498 &  20.223 &   0.017  \\ 
cep001 & V  &  2932.641846 &  20.288 &   0.015  \\ 
cep001 & V  &  2939.582520 &  20.501 &   0.024  \\ 
cep001 & V  &  2946.688965 &  20.655 &   0.043  \\ 
cep001 & V  &  2952.562744 &  20.776 &   0.028  \\ 
\enddata

\end{deluxetable}

\begin{figure}[htb] 
\vspace*{18cm}
\includegraphics{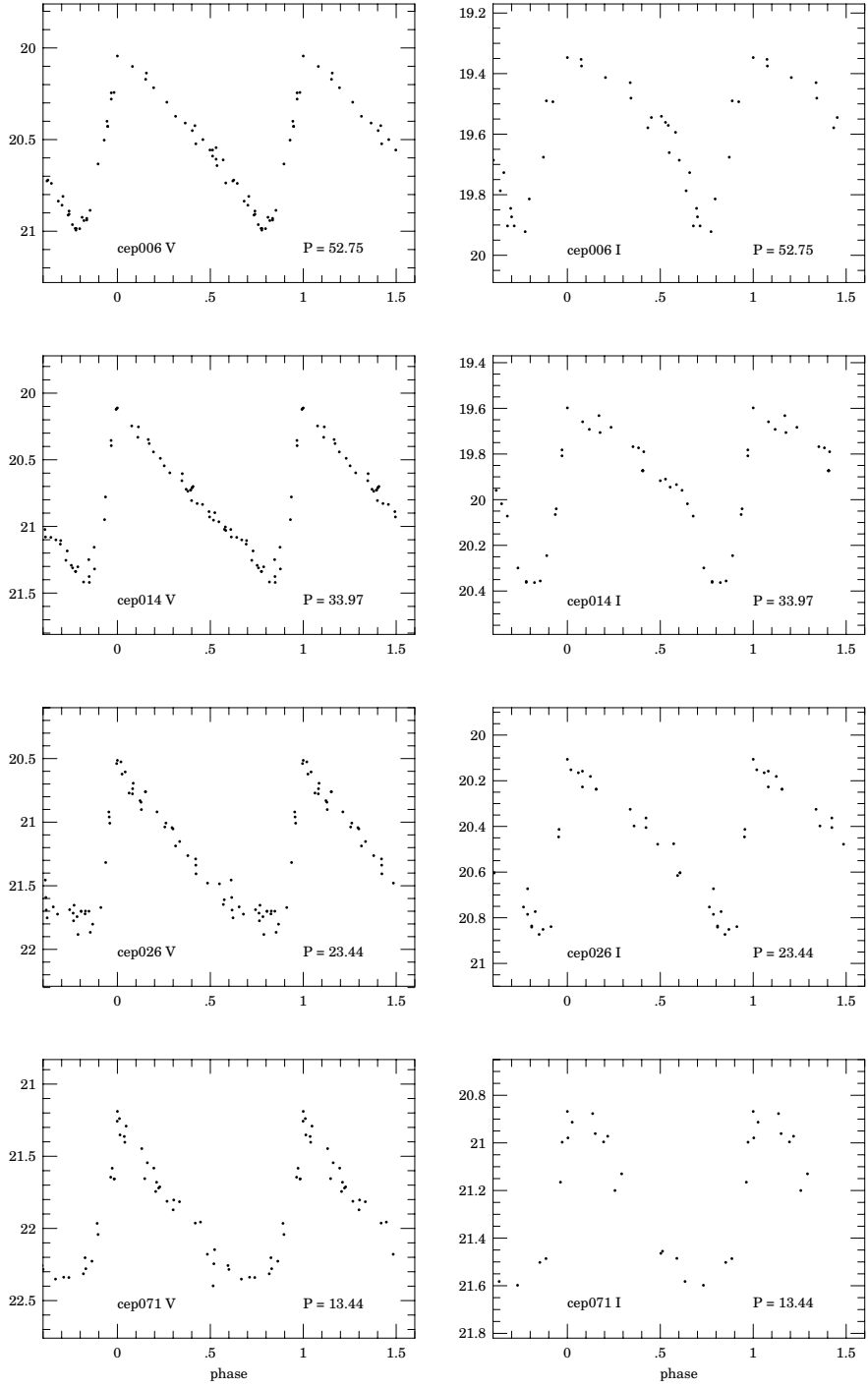} 
\caption{Phased V- and I-band light curves for Cepheids of different periods 
in our NGC 300 sample.
}
\end{figure}  

\begin{figure}[htb]
\vspace*{20cm}
\includegraphics{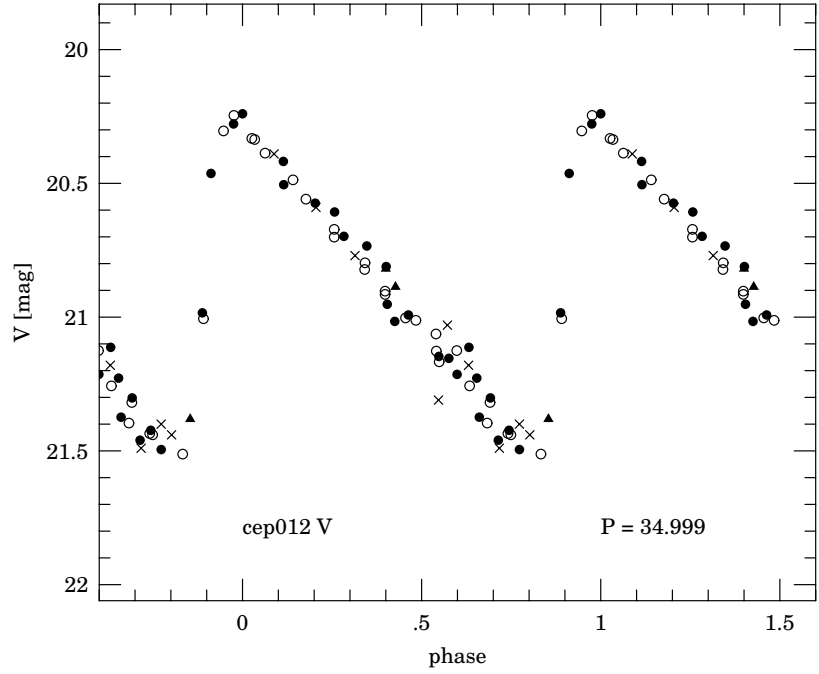}
\includegraphics{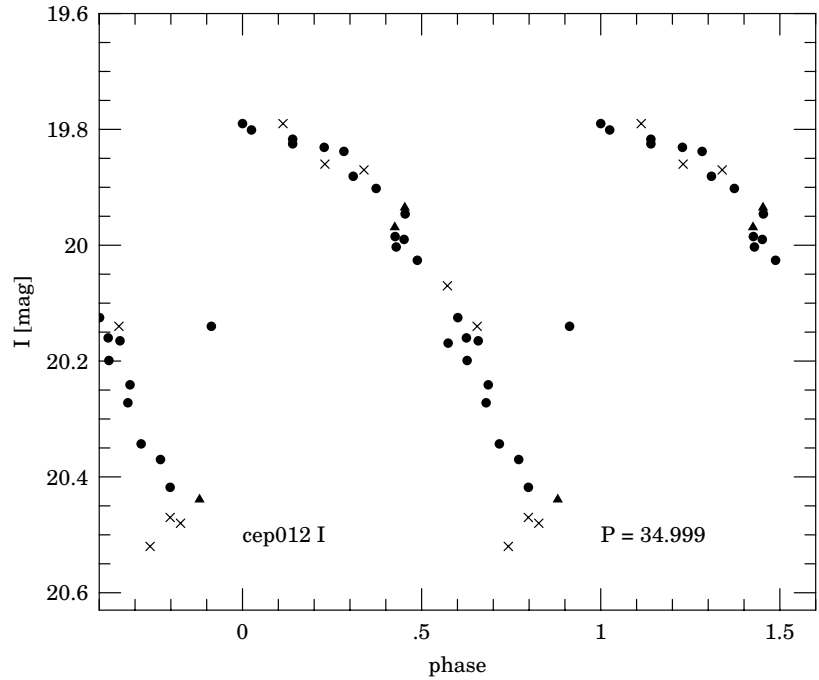}
\caption{The phased V- and I-band light curves of the NGC 300 Cepheid cep012.
Filled circles, our new Warsaw 1.3 m data. Open cicles, our previous 
ESO/WFI data. Filled triangles, our new CTIO 4 m observations. Crosses, 
data from Freedman et al. (1992). The good agreement of the different 
datasets is evident.
}
\end{figure}

\begin{figure}[htb]
\vspace*{15cm}
\includegraphics{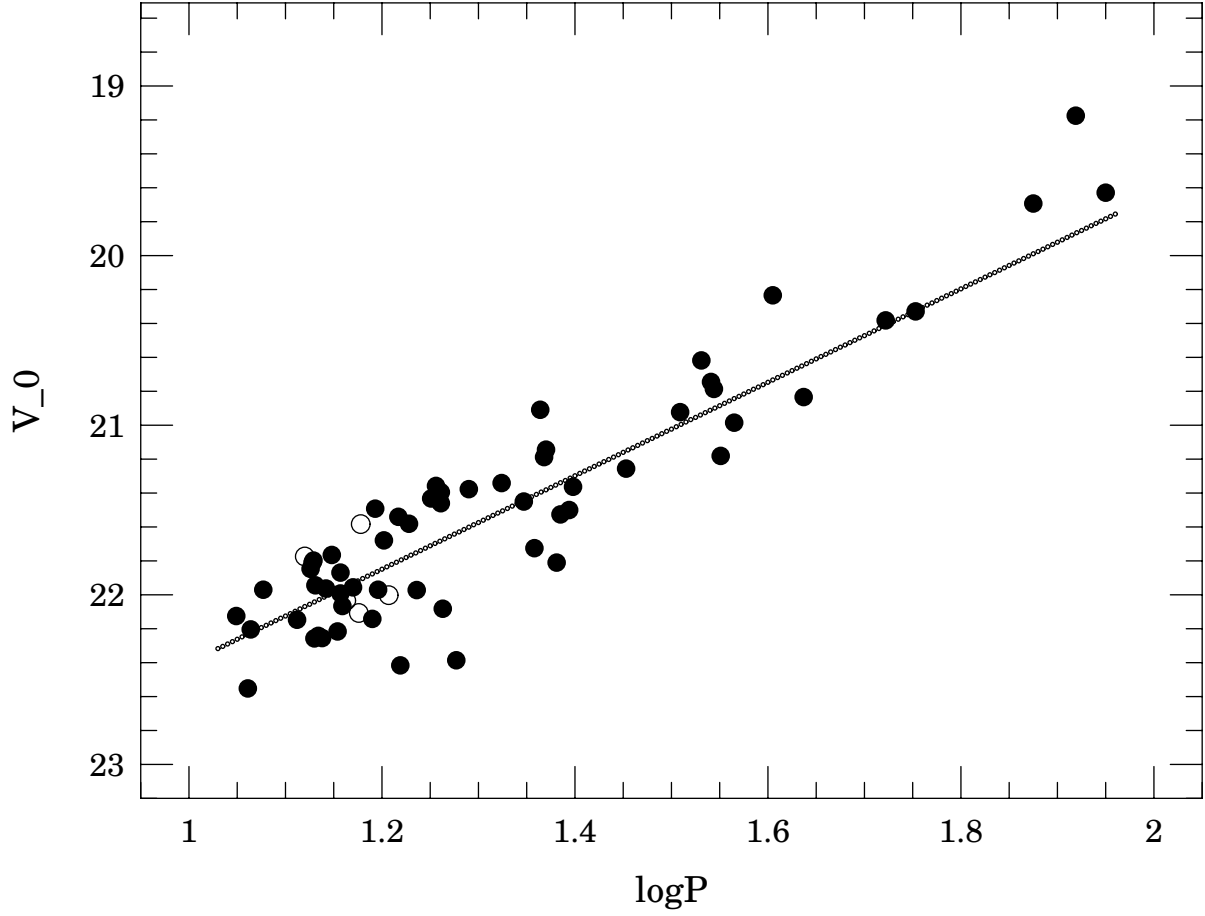}
\caption{The period-luminosity relation for NGC 300 Cepheids in the V 
band. Open circles denote variables not used in the distance determination.
The slope of the relation was adopted from the LMC Cepheids (OGLE II).
}
\end{figure}

\begin{figure}[htb]
\vspace*{15cm}
\includegraphics{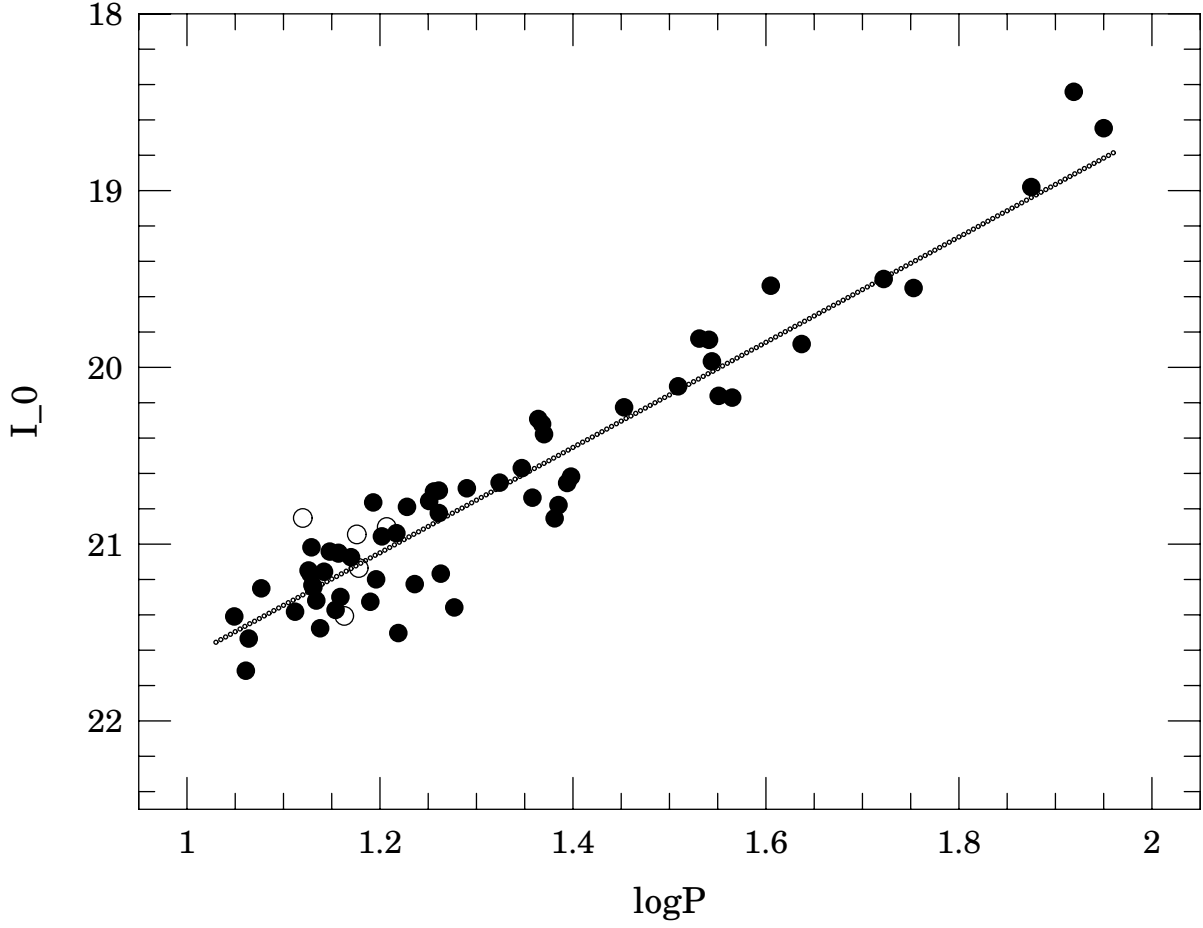}
\caption{Same as Fig. 3, for the I band.
}
\end{figure}

\begin{figure}[htb]
\vspace*{15cm}
\includegraphics{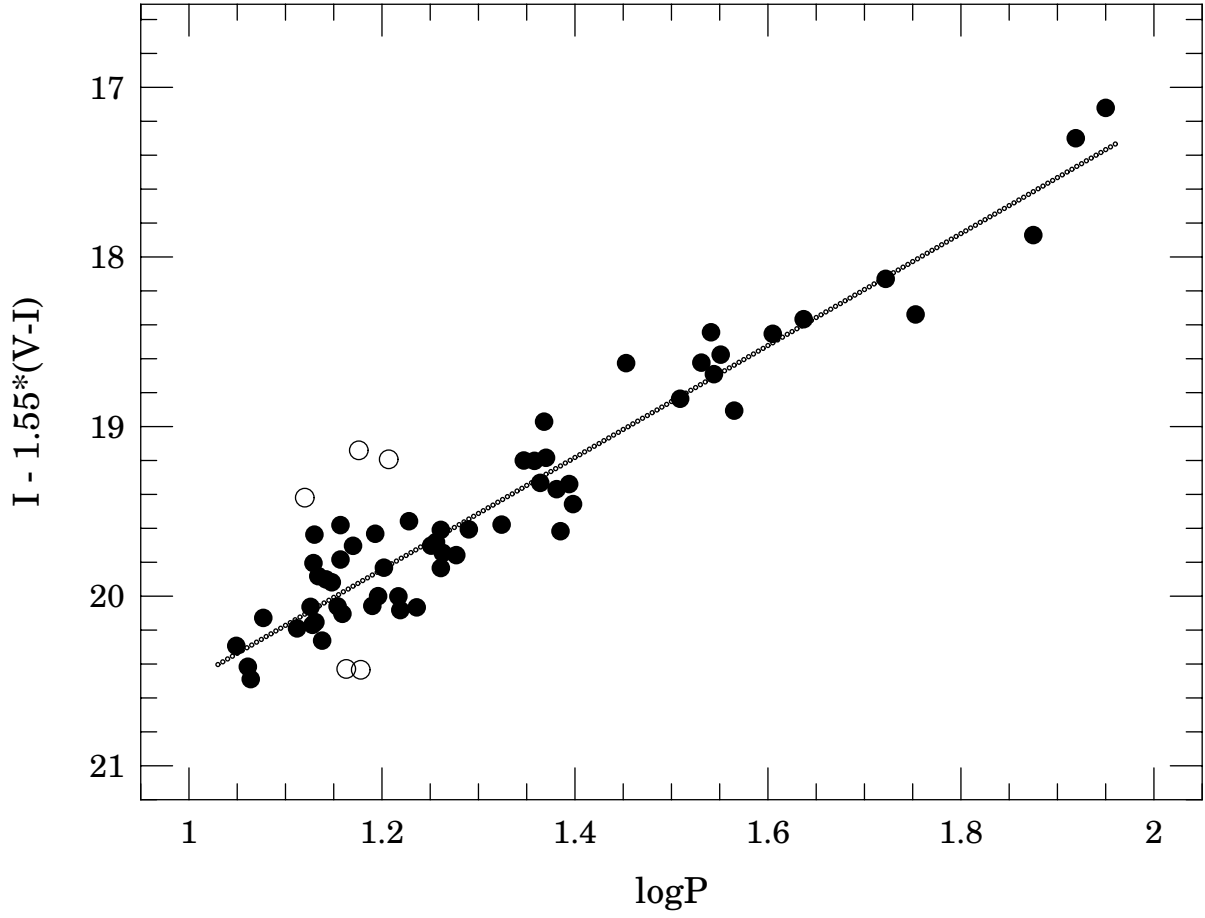}
\caption{Same as Fig. 3, for the reddening-independent (V-I) Wesenheit 
magnitudes.
}
\end{figure}

\end{document}